\documentclass[manuscript,natbib=false]{acmart}
\usepackage{graphicx} % Required for inserting images
\usepackage{enumitem}
\usepackage{acronym}
\usepackage{xcolor}
\usepackage{siunitx}
\usepackage{tikz}

\usetikzlibrary{shapes,arrows}
\usetikzlibrary{positioning}
\acrodef{RQ}{Research Question}
\acrodef{RtD}{Research-through-Design}
\acrodef{NCA}{National Competent Authority}
\acrodefplural{NCA}[NCAs]{National Competent Authorities}
\acrodef{AIRS}{AI Regulatory Sandbox}
\acrodefplural{AIRS}[AIRSes]{AI Regulatory Sandboxes}
\acrodef{AITS}{AI Technical Sandbox}
\acrodefplural{AITS}[AITSes]{AI Technical Sandboxes}
\acrodef{BNA}{Boundary Negotiating Artifact}
\acrodef{EU}{European Union}
\acrodef{EU AI Act}{EU Artificial Intelligence Act}

\title{Designing a Boundary Negotiating Artifact for Collaborative Socio-Technical Sense-Making in AI Regulatory Sandboxes}
\author{Idoia Landa-Oregi}
\affiliation{%
  \institution{Luxembourg Institute of Science and Technology (LIST)}
  \city{Esch-sur-Alzette}
  \country{Luxembourg}
}
\email{idoia.landa@list.lu}

\author{Tom Deckenbrunnen}
\affiliation{%
  \institution{LIST, University of Luxembourg}
  \city{Esch-sur-Alzette}
  \country{Luxembourg}
}
\email{tom.deckenbrunnen@list.lu}

\author{Alessio Buscemi}
\affiliation{%
  \institution{LIST}
  \city{Esch-sur-Alzette}
  \country{Luxembourg}
}
\email{alessio.buscemi@list.lu}

\author{Daniele Pagani}
\affiliation{%
  \institution{LIST}
  \city{Esch-sur-Alzette}
  \country{Luxembourg}
}
\email{daniele.pagani@list.lu}

\author{German Castignani}
\affiliation{%
  \institution{LIST}
  \city{Esch-sur-Alzette}
  \country{Luxembourg}
}
\email{german.castignani@list.lu}
\date{}

\RequirePackage[
datamodel=acmdatamodel,
style=acmnumeric, % use style=acmauthoryear for publications that require it
]{biblatex}

\begin{document}

\begin{abstract}
    The rapid, unpredictable advancements in AI system capabilities has seen regulators take adaptive and experimental approaches to policymaking.
    Established in other domains as instruments balancing regulation with innovation, regulatory sandboxes are seen as solutions for AI regulation.
    However, analyses mostly focus on the legal and institutional design of \acp{AIRS}.
    With the legal framework leaving the socio-technical interpretation to stakeholders, this creates a gap on the sense-making required to fulfill the AIRS purpose.
    In this paper, we approach this by designing a Boundary Negotiating Artifact as a way to mediate meaning in \acp{AIRS}.
    Through Research-through-Design we iteratively develop a tool, providing an interface for the different stakeholders to collaborate in AI assessment. We then position it as technical backbone in established AIRS frameworks, structuring the collaborative sense-making of the involved stakeholders.
    We further report the insights gained from our design process leaving the qualitative evaluation for future work.
\end{abstract}

\maketitle
\acresetall

\section{Introduction}
\label{sec:intro}
An \ac{AIRS}, as defined by Articles~57 and 58 of the \ac{EU AI Act}, is a controlled environment intended to foster innovation across the AI system life cycle and contribute to regulatory learning~\cite{EUAIAct2024}.
It consists of the supervision of a prospective AI system provider by a \ac{NCA}, in particular through legal guidance, but also through technical expertise and resources.
Thus, the provisions laid out by the \ac{EU AI Act} create the institutional space for collaboration between diverse stakeholder groups from legal to technical communities. Beyond the technical or legal challenges these stakeholders might face, a big challenge lies in the need for sense-making \cite{guioespanolRegulatorySandboxesAI2025}. How do stakeholders with different expertise, backgrounds, vocabularies or interests build a shared understanding to make consequential decisions together under the uncertainty of AI regulation?

This is fundamentally a collaborative sense-making problem, and one that the HCI discipline is well-positioned to address. Being at the intersection of inter disciplinary communities, \acp{AIRS} generate information asymmetries, referring to the uneven access or understanding of relevant information among stakeholders, which can affect the transparency and decision-making of the collaborative process. Nevertheless, with the appropriate artifacts, \acp{AIRS} can work as levers to reducing said information asymmetries, pairing designed testing mechanisms with the combined expertise of the stakeholders involved \cite{Espanol2025RegulatoryAdaptation}. However, regardless of the various efforts on \ac{AIRS} definition and implementation, little is known about how this information asymmetry is actually experienced by the people who must navigate it, or how it might be addressed in practice. 

Consider what a sandbox encounter requires for its stakeholders. An AI innovator must translate a system they have developed and understand in technical terms into a language of compliance, with the conformity obligations that the regulatory framework demands. At the same time, a regulator must assess a system they have never encountered before, in a specific sector, using legal provisions that were defined in a level of abstraction that does not clearly map to technical specifications. Both stakeholders carry genuine expertise of their own field, however, they lack the vocabulary of the other. There is a gap in communication that the regulation does not address. \acp{AIRS} are defined as a legal process, offering a sound procedure but without the sense-making infrastructure that can make the process adequate to all stakeholders.

Prior work has analyzed \acp{AIRS} as institutional environments for multi-stakeholder boundary negotiation over AI systems and AI regulation~\cite{Deckenbrunnen2026BathtubsUncertainty}. Particular focus was put on the role of \acp{BNA} as effective support of collaborative sense-making and boundary object stabilization, enabled by the legal uncertainty caused by abstract legislation. A \ac{BNA}, being documentation, a model, or a process, can be used by any stakeholder group to communicate their perspective to another group, reducing the information asymmetry. While using a \ac{BNA} can genuinely reduce the information asymmetry, its design strongly relies on a deep understanding of its prospective users. Understanding their current experience with AI regulations and testing, pain points, and needs is essential for the successful definition of an artifact that will ease their overall experience. Applied to \acp{AIRS}, this means that a \ac{BNA} cannot be purely designed from a legal text, nor for a technical best practice alone, as it should be shaped by how innovators and regulators actually experience the sandbox encounter, the assets they each bring, and the points of divergence in their understanding. This stakeholder-centered view is what allows \ac{BNA} to move from a compliance artifact to a genuine sense-making tool.

This paper addresses this gap by designing a tool that works as a backbone for structuring the collaboration of stakeholders in an \acp{AIRS}, reducing the information asymmetry. Adopting a \ac{RtD} approach, it first understands the experience of regulators and innovators in this context and then designs a tool to enhance this collaboration and expand the consensus \ac{AIRS} framework with a technical \ac{BNA}.

Within this context, this paper reports a \ac{RtD} investigation conducted through qualitative research with AI innovators, technical experts, and regulatory authorities, and through an iterative design process producing a tool to mediate their interaction. Specifically, the research is guided by the following \acp{RQ}:

\begin{enumerate}[label={$\bullet$~\textbf{RQ\arabic*:}}, left=\parindent]
    \item How do AI innovators, technical experts and regulatory authorities differently understand, experience, and navigate the obligations and ambiguities of AI regulations?
    \item What design knowledge emerges from creating a tool that mediates that experience?
    \item How can that knowledge inform \ac{AIRS} operationalization?
\end{enumerate}

\section{Research context}
\label{sec:background}

\subsection{Legal and Institutional Design of AIRS in Europe}
\label{sec:AIRS-legal}

\acp{AIRS} were included in the \ac{EU AI Act} as a mechanism to support innovation and regulatory learning under conditions of legal uncertainty, especially for smaller organizations that often lack the resources to navigate compliance on their own. Within a regulatory sandbox, a \ac{NCA} provides an AI innovator with direct regulatory guidance and supervisory support. This provision is inspired by established precedent from financial regulations, where sandboxes have been used since 2016 to let fintech innovators test novel products under supervised conditions, an approach that has since been adopted by many other sectors\cite{Fahy2022FosteringFintech}. 

The \ac{EU AI Act} translates this into an institutional structure, requiring every member state to establish at least one national \ac{AIRS} operated by the designated \ac{NCA}. In parallel, the AI office is responsible for a sandbox addressing general-purpose AI systems at \ac{EU} level. Beyond this structure, however, the \ac{EU AI Act} leaves considerable room for variation, as \acp{AIRS} may differ by sector, technical expertise of the \ac{NCA}, or the specific testing infrastructure they are able to offer. For instance, an AI innovator operating across multiple member states or a system intended for cross-boarder deployment may need to satisfy expectations that differ from one national sandbox to the next. This fragmentation in itself is a motivation for the framework proposed in this paper, as a shared tool and process structure offering a single path toward the kind of interoperability that current provisions lack.

Several scholars have examined how \acp{AIRS} should be designed and governed. \citeauthor{novelliGettingRegulatorySandboxes2026} provide an in-depth description of \ac{AIRS} design from a regulatory lens, detailing the steps and tasks that structure each step of the process. \cite{Chen2026DesigningObjectives} complement this work with a framework of characteristics, design choices, and functionalities an \ac{AIRS} needs to fulfill its intended purpose. Similarly, \cite{Espanol2025RegulatoryAdaptation} analyze best practices for existing sandboxes to support a more robust inquiry into the learning process within an \ac{AIRS}. This body of work establishes \textit{what} a well-informed \ac{AIRS} should consider both procedurally and structurally. What remains underexplored is \textit{how} the stakeholders engaging in such an environment practically communicate and build shared understanding.

 Following these guidelines, some \ac{EU} Member States have made concrete progress toward implementing their \ac{EU AI Act} mandated sandboxes, but the considerable discretion left to \acp{NCA} risks producing fragmented or incompatible approaches that could undermine the interpretation of the \ac{EU AI Act} \cite{Ahern2025OperationalisingProviders}. This concern is echoed at the institutional level as the European Parliament has flagged that fragmented enforcement could leave national authorities with uneven capacities and providers intentionally selecting less strict sandboxes as a result \cite{EuropeanParliment2026AIParliament}. The \ac{EU AI Act} itself anticipates this risk, tasking the Commission with adopting implementing acts under Article 58(2) to establish common principles on eligibility, application, and exit procedures precisely "in order to avoid fragmentation across the Union" \cite{EUAIAct2024}. However, this harmonization strategy operates at the level of legal procedure and does not address the day-to-day, cross-stakeholder experience of navigating a sandbox.

 \subsection{Boundary Negotiation in Multidisciplinary Collaborative Work}
 \label{sec:boundary-negotiation}

Understanding the complex interactions arising in collaborative work between diverse stakeholder groups is a non-trivial task.
Through a series of ethnographic studies examining such interactions, \citeauthor{starInstitutionalEcology1989} noted that even in contexts with no consensus among communities of practice, the collaboration tends to be structured around objects with a specific set of properties~\cite{starInstitutionalEcology1989}.
Such objects, be they material or procedural, are robust enough to serve as a shared core between stakeholders and, at the same time, possess the interpretive flexibility to accommodate the diverging information needs of different groups.
\citeauthor{leighstarThisNotBoundary2010} expands on her previous work through an account of the life cycle of boundary objects~\cite{leighstarThisNotBoundary2010}.
The creation and stabilization of boundary objects rely on phases attempting to standardize ill- and well-structured interactions among stakeholders, and the introduction of new perspectives with engagement of new communities of practice.
\citeauthor{leighstarThisNotBoundary2010} further raised concerns about the misuse of the concept of boundary objects as any object offering any amount of interpretive flexibility.
The analysis by \citeauthor{leeBoundaryNegotiatingArtifacts2007} notes a disconnect between the boundary object concept and the methods of standardization with the concept becoming a catch-all a highlighted issue~\cite{leeBoundaryNegotiatingArtifacts2007}.
The paper addresses this disconnect by introducing the notion of boundary negotiating artifacts, artifacts that are distinct from boundary objects and serve as a support in collaboratively negotiating the meaning and eventually stabilizing the boundary objects in question.
More work has investigated the use and design of \acp{BNA} in interdisciplinary, collaborative work in both engineering and knowledge work contexts~\cite{beddoesUsingBoundaryNegotiating2011a, leeEmbracingChaosAgain2026}.
Beyond these domains, boundary objects and \acp{BNA} have also been used to harmonize different vocabularies and methods across product design practices \cite{Velleu2023AMETHODS}, to coordinate distributed roles and responsibilities in agile systems \cite{Wohlrab2019BoundaryEngineering}, and to bridge epistemological differences between qualitative and quantitative traditions in mixed-methods research \cite{Wenger-Trayner2019BoundariesResearch}. Together, this work illustrates how boundary objects and \acp{BNA} can work across disciplines wherever heterogeneous expertise must be harmonized.

The AI compliance audit space, due to rapid technological advances and relative regulatory lethargy, is marked by a lack of established standardized practices~\cite{schiffEmergenceArtificialIntelligence2024}.
However, as \citeauthor{schiffEmergenceArtificialIntelligence2024} points out, it is the work of auditors and developers that leads to new advancements in auditing methodologies and practices.
\acp{AIRS} in themselves are institutionalized environments that bring together technical and regulatory stakeholders.
At their core, they are intended to serve in support of innovation and towards reducing legal uncertainty caused by abstract legislation.
This uncertainty further complicates the inter- and multidisciplinary collaboration.
While it may offer the space for productive cooperative sense-making and the standardization of practices, it requires appropriate \acp{BNA} for the process to be effective~\cite{Deckenbrunnen2026BathtubsUncertainty}.
\citeauthor{Deckenbrunnen2026BathtubsUncertainty} therefore interpret \acp{AIRS} as boundary negotiation environments with a given AI system and applicable AI regulation as its boundary objects.
The authors highlight components and requirements of technical frameworks for them to serve as \acp{BNA} in an \ac{AIRS}, such as collaborative dashboards and role-specific access.

\subsection{Human Experience of Stakeholders in AIRS}
\label{sec:human-exp-ai-assessment}

The proper design of a \ac{BNA} to bridge the information asymmetry among stakeholders relies on the understanding of their experience and needs.
While offering services in AI assessment our research team has engaged with multiple industry partners which has given us meaningful insights into stakeholder needs.
These insights summarize the current experience and needs of the three key stakeholders,  AI innovators, regulators, and technical experts.
The AI innovator is a company or organization developing an AI system that needs to be tested. Technical experts, are third party assessment experts who are familiar with AI legal requirements and have the know-how to perform technical assessment of AI systems. Lastly, the \acp{NCA} are the ones that need to help with the interpretation of results and give guidance and support to AI innovators. 

\textbf{AI innovators}. AI innovators are experts in building AI systems, but may lack the expertise or resources to conduct the assessment of their system, beyond more immediate performance metrics. AI regulations, such as the \ac{EU AI Act}, present high regulatory complexity to AI innovators, introducing many requirements that are difficult to interpret without assessment or legal expertise. Knowing what test to use for their specific use case, finding said test, learning how to perform the test, or interpreting the results and compiling the results in a report are activities that most innovators struggle with. Moreover, these regulations include extensive documentation requirements, including the production of evidence and the maintenance of technical documentation throughout the AI life-cycle.

AI innovators face the challenge of balancing innovation activities with regulatory obligations, as they often perceive that compliance activities increase development time and administrative burden, making it more difficult to iterate rapidly while remaining competitive. Moreover, they find difficulties in translating legal requirements into technical practices, as concepts are expressed in legal language that is not always easily translated into engineering tasks. Lastly, the evolving guidance poses a challenge. Although the \ac{EU AI Act} establishes the legal framework, many implementation details are still to be addressed by standards, guidelines, and codes of practice, with their current absence creating legal uncertainty during development.

\textbf{Technical experts}. A prominent way of conducting AI assessment is through digital sandboxes \cite{Buscemi2026OperationalisingAct, eusair_roadmap_2025}. However, the set up of technical environments for AI assessment remains highly complex and time-consuming, creating a significant barrier to adoption at scale. Technical experts have to be able to identify relevant requirements and translate those into metrics and tests. After that, they have to find the relevant tests from a wide variety of sources and manually integrate them to execute the tests. Once the testing is done, they have to manually harmonize the results of each test into a coherent database and dashboard and manually collect the evidence of the tests that were performed. Lastly, they have to interpret the results with feedback from different experts and manually compile everything into a report.

There is a growing demand for AI assessment in the \ac{EU} that current assessment practices cannot fulfill. The setup of the \ac{AITS} is unique to each use case or intended purpose, which requires extended time and effort. Each use case and sector has different implications, with specific risks, metrics, testing methods and thresholds, where a common \ac{AITS} is not viable. Lastly, there is a clear need for multi-disciplinary collaboration. 

\textbf{\acp{NCA}}.
While the AI Act clearly sets out the responsibilities of \acp{NCA} regarding the AIRS, the operationalization of these sandboxes remains unclear, leaving significant discretion regarding their governance, criteria, testing procedures, and evaluation methods. Authorities must encourage experimentation while ensuring compliance, a balance that is difficult to achieve in practice. Moreover, most AI applications often fall under several regulatory domains, requiring coordination among multiple regulatory stakeholders with different expertise.

\acp{NCA} have limited to no technical expertise in AI assessment depending on the country, making it difficult to effectively supervise innovative AI systems. While they provide the regulatory expertise, they need practical methods and guidance to evaluate compliance with requirements such as robustness, transparency, or human oversight.

While the technical experts are the ones setting up the environment for assessing the AI system, the interpretation of the produced results requires cooperation among different experts and entities.
In fact, contextualization in the given use case is essential for proper interpretation of the assessment results.
For instance, an AI system for object detection, achieving $\SI{99}{\%}$ accuracy, may be appropriate for a security camera with an alert system, but wholly unacceptable for pedestrian detection in an autonomous car.
Similarly, abstract notions such as bias or other risks to fundamental rights are difficult to capture in a technical, one-size-fits-all manner, as established technical metrics may be met, but ignore, e.g., legal implications arising from case law.
This makes collaborative, multi-stakeholder sense-making a particularly integral part of AI assessment.

\section{Methods}
This study follows a \ac{RtD} approach, where the research is guided through the design process logic and supported by scientific research questions \cite{Zimmerman2007ResearchHCI}. The design process involves various iterations, favoring a constant realignment of the artifacts to better tackle the design problems \cite{Cortesao2022ResearchPractice,Godin2014AspectsReview}. The method is applied to develop a framework that facilitates the interactions of stakeholders in an \ac{AIRS}, built upon the findings gathered from the design and development of a tool to facilitate the configuration of \acp{AITS}.

\begin{figure}[!htb]
    \centering
    \includegraphics[width=1\linewidth]{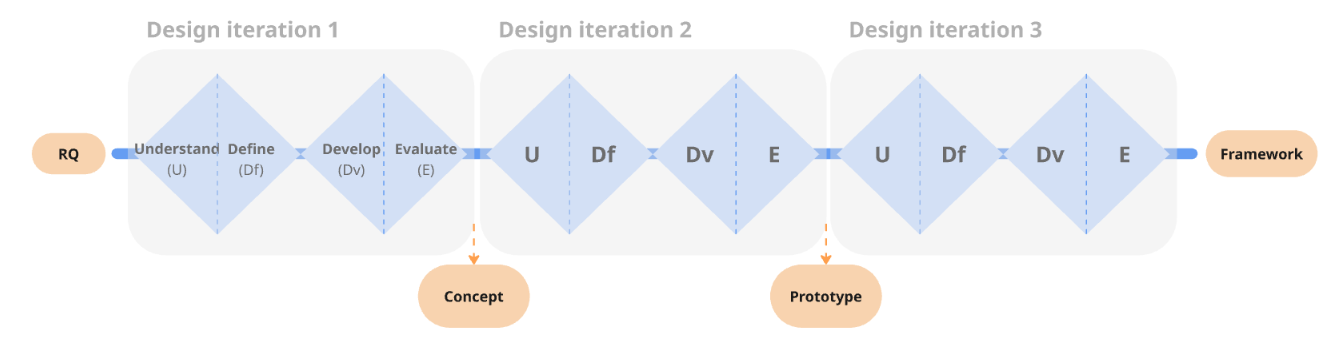}
    \Description{Design process followed in the study, starting from the research questions and followed by 3 design iterations, each with the double diamond, concluding in the framework}
    \caption{Design process}
    \label{fig:prozesua}
\end{figure}

The \ac{RtD} process, shown in Figure \ref{fig:prozesua} is guided by the three \acp{RQ} defined in Section~\ref{sec:intro}.
With the context of \acp{AIRS} in the \ac{EU} and the the current experience of the different stakeholders as a starting point, we developed an open source tool (whose name is hidden due to double blind review, but whose identity and url will be included upon acceptance of the manuscript) to mediate the experience and interactions of the stakeholders through three design iterations. 
The design process for the development of the tool followed the main phases of Human-Centered Design \cite{Giacomin2014WhatDesign} through three design iterations (see Figure \ref{fig:prozesua}). The first phase, \textit{understand}, focuses on developing a grounded understanding of the problem, user needs, and the overall design context. The second phase, \textit{define}, focuses on synthesizing all the insights gathered in the first phase to frame the challenge and requirements for the design. The third phase, \textit{develop}, consists in generating ideas and developing potential solutions. In the last phase, \textit{evaluate}, those solutions are tested with the end users to gather feedback and refine the product through iterative loops. The main value of this approach lies in engaging users and stakeholders in each of the iterations to address user needs while still maintaining its viability \cite{Brown2008DesignThinking}.

With every design iteration the expectations of the different stakeholders were adjusted, making sure the product addressed their challenges and clarified the roles and responsibilities in an \ac{AIRS}. 
The tool was used as an instrument for producing knowledge throughout the design process, as it served as a baseline for developing the framework presented in Section~\ref{sec:framework}. This framework presents how the three main stakeholders potentially interact in an \ac{AIRS} having the designed tool as a main touchpoint. It highlights the roles of each stakeholder and their exchanges through the tool.

\section{Designing a computer-assisted BNA for AIRS} 
\label{sec:design-tool}
This section presents the findings gathered from the design of a digital tool that facilitates the exchanges among stakeholders by helping users in the configuration of \acp{AITS} to assess AI systems. These findings served as a baseline to develop the framework presented in Section \ref{sec:framework} that outlines the interactions among the stakeholders in the operationalization of \acp{AIRS}. Thus, this section reports on the details of the three design iterations in the development of the tool, giving special focus to the associated learnings.

The tool provides a curated, extensible catalog of AI tests, controls and datasets integrated through an open plug-in API interface. It guides users through mapping regulatory requirements to assessment metrics, selecting from the catalog the tests and controls relevant to their specific use case, configuring a collaborative dashboard to execute tests and visualize results, and generating structured assessment reports. 

\subsection{Design Conception}
\label{sec:concept}
The first design iteration had the purpose of developing a concept that would bridge the different experiences of the users into a single harmonized platform. Without going into the development of the digital platform yet, this cycle consisted in testing the concept with the relevant stakeholders. In order to do so, the concept was first proposed in a white paper by Buscemi et al. \cite{Buscemi2026OperationalisingAct} presenting the general concept and architecture. 

To validate the concept, a presentation was made by attending Nexus, one of the premier technology events accelerating AI and technological solutions in the \ac{EU}. This is a multidisciplinary event gathering attendants of all the relevant stakeholders mentioned in section \ref{sec:human-exp-ai-assessment}. Therefore, this approach allowed the team to gather insights from all the relevant stakeholders at once. The concept was informally presented to ten attendees, from all three stakeholder groups. The feedback was gathered through a semi-structured discussion and analyzed by the team through debrief notes to identify possible improvements. 

With the \ac{EU AI Act}'s general provisions and prohibitions applying shortly before the conference, the concept was well received especially among the technical stakeholders, understanding it as a bridge between the technical and regulatory needs. The possibility to have a tool that has the potential to help them navigate all the needed testing of their systems, in a harmonized way, was highly appreciated. However, only technical users really understood the concept at this point, with the original design proving too technical for easy understanding by non-technical stakeholders. This highlighted the need to establish a less technical path in the tool, where regulators can access the results of the testing in a simplified manner and to be able to track the testing that was conducted.  

These findings shaped the first design of the digital tool (presented in Section \ref{sec:demo}) and helped formalize the interactions of the stakeholders in the core phases of the framework, \textit{testing} and \textit{evaluation}, presented in Section \ref{sec:framework}.

\subsection{Building a Demonstrator}
\label{sec:demo}
The second design iteration consisted in building the first prototype of the digital tool, based on the concept validated in the first iteration and refined with the insights gathered from potential users. This involved developing a catalog that integrates a wide variety of tests, a test execution engine to run those tests in a harmonized way, dashboards to present the results to the different stakeholders, and a report generator. This prototype was developed and tested in the context of \ac{EU} regulatory sandboxes, which served as the primary case for grounding the tool's requirements against a live regulatory process.

\begin{figure}[!htb]
    \centering
    \includegraphics[width=1\linewidth]{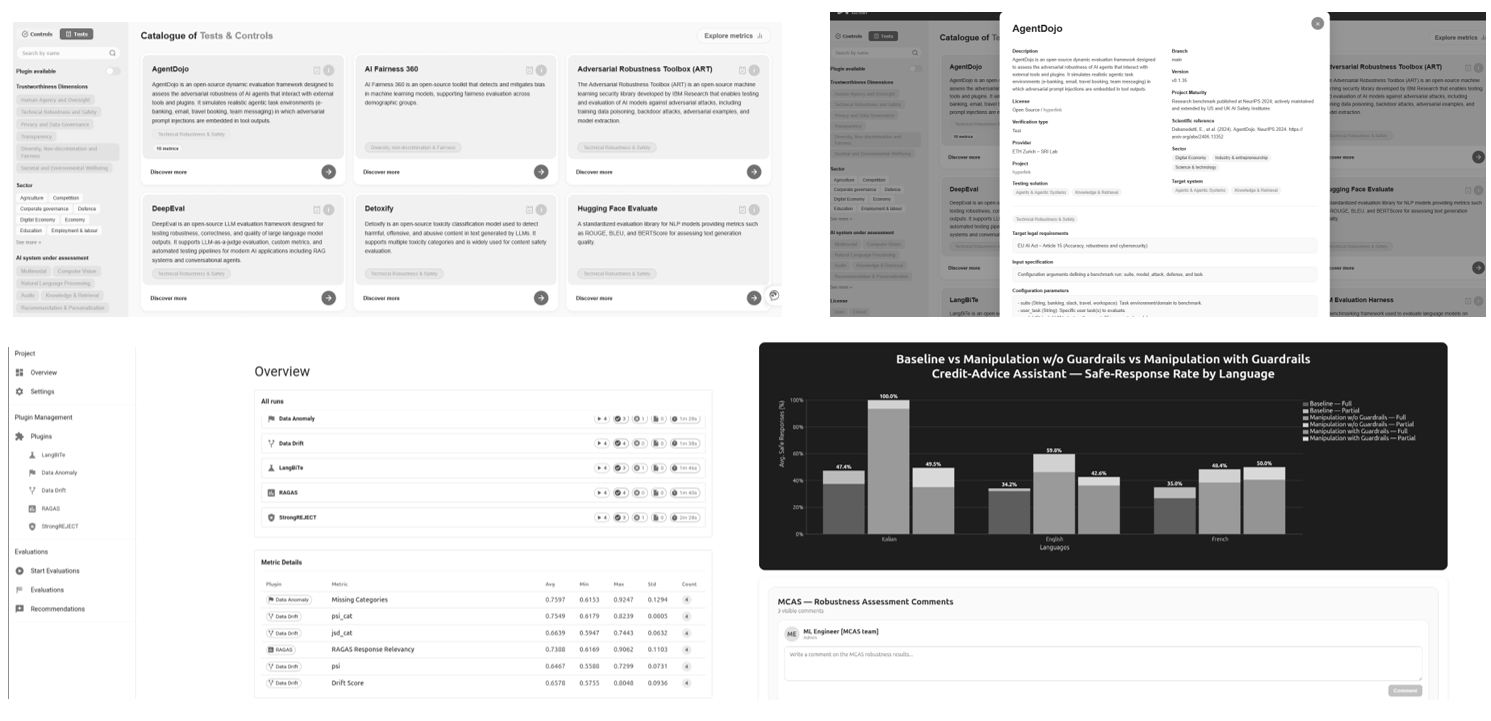}
    \Description{wireframes of the 4 main screens of the tool: catalog of tests, harmonized test description, test execution engine, dashboard}
    \caption{Wireframes of the main screens: catalog of tests, test execution engine and dashboard}
    \label{fig:general w}
\end{figure}

The demonstrator was evaluated in two parallel settings to gather feedback from the different stakeholder groups. First, it was presented at World AI Cannes Festival, one of the leading \ac{EU} events on artificial intelligence, where it was presented informally to 40 attendees, especially AI innovators and technical experts. Then, it was presented in a focused workshop with 60 participants, gathering \acp{NCA} and representatives from different \ac{EU} initiatives, such as AI Factories, Digital Innovation Hubs and Testing and Experimenting Facilities. This dual approach allowed to have a more targeted feedback, understanding the acceptance of the different stakeholders towards the tool and the remaining issues to be improved. The feedback was gathered through an observed walkthrough of the demonstrator followed by a semi-structured discussion and analyzed by the research team through debrief notes to identify recurring pain points.

The main insight gathered from this evaluation was that the tool was very focused on the technical assessment and technical users. Not all organizations that will participate in an \ac{AIRS} will have to do technical testing, depending on the risk level of the AI system under assessment they might only need to complete lighter controls. Moreover, several participants highlighted that not all users will be able to qualify their system on their own, being a relevant feature that can be added to the tool, improving the overall user experience, mostly in the starting phases of the assessment.

These findings shaped the decisions carried out in the third iteration (Section \ref{sec:moving}) and helped formalize the qualification step in the on-boarding phase of the framework (Section \ref{sec:framework}) by introducing an initial system qualification step ahead of the test selection and broadening the scope of the tools to integrate controls for lower risk systems.

\subsection{Moving Beyond the Technical}
\label{sec:moving}
\begin{figure}[!htb]
    \centering
    \includegraphics[width=0.75\linewidth]{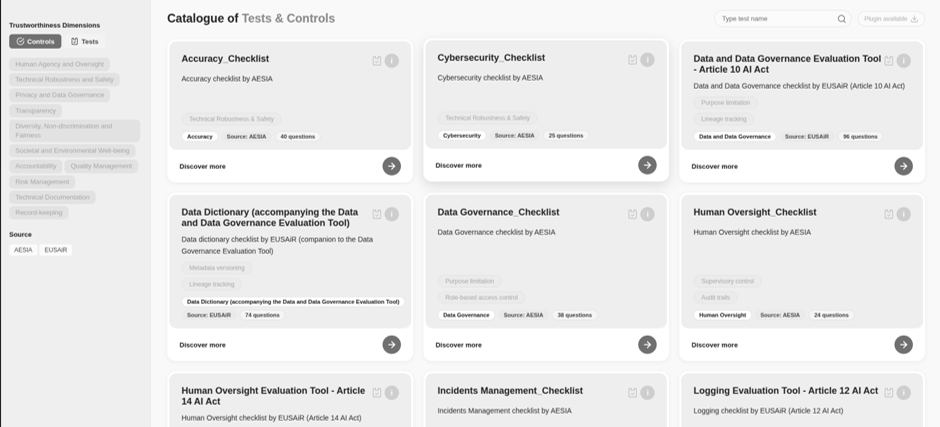}
    \Description{Wireframe of the catalog, including the control update, complementing the tests}
    \caption{Wireframe of the catalog including controls}
    \label{fig:catalog}
\end{figure}

The last iteration presented a more holistic approach, involving non-technical users in a more meaningful way, as the core innovation of the tool is not only its technical capacity, but the ability to provide a common ground for all stakeholders regarding AI assessment. The first prototype was improved by integrating aspects such as an initial prototype for the qualification of the AI system and control checklists into the catalog. For the AI system qualification, the user will define in detail the system they want to assess, such as intended purpose, target users, and high-level information about data governance, human oversight, transparency to end users. It also includes the selection of several tags such as to identify the types of AI technologies on which the system is based and the sectors of application. The catalog includes control checklists from diverse sources~(see \ref{fig:catalog}). Similarly to the tests, each control checklist is represented by its slug, which can be expanded to retrieve information such as source, addressed regulation, or standard addressed or specific obligations.

The improved prototype was presented at the following edition of the first conference (Section \ref{sec:concept}), ensuring attendance of all relevant stakeholders. Along with the formal presentation at this event, the prototype was also released as an open-source tool for technical users to test. It was formally presented in the conference to 30 attendees in a focused session and the demonstration was available for informal presentation during the two days of the conference, reaching another 40 attendees for more detailed one on one discussions. All main stakeholders were represented among the attendees, providing balanced feedback across all groups. As in the previous iteration, the feedback was gathered through an observed walkthrough of the demonstrator followed by a semi-structured discussion and analyzed by the research team through debrief notes to identify recurring pain points. 

The prototype was well received by the ecosystem, both by technical and non-technical users. Overall, stakeholders appreciated a platform that helped them navigate the wide variety of testing options and existing controls in a simplified way and the capacity to share the results with the relevant stakeholders through the collaborative dashboard and exit report. Thus, the harmonization of the different tests was also highly appreciated by the ecosystem, as the experience of the users was simplified to learning how to perform the test once, instead of starting from scratch with every new test.
These findings shaped the final details of the framework (Section \ref{sec:framework}), detailing the interactions among technical users and regulators through the use of the tool. They helped shape further all actions in the different phases of the framework, filling the gaps of the journey of the three stakeholders.

\section{A Technical BNA as Backbone in AIRS Frameworks}
\label{sec:framework}
The successful implementation of \acp{AIRS} requires a level of collaboration that challenges the operationalization of said environments, as the different stakeholders have different needs and inputs throughout the assessment process. The findings from understanding the stakeholders, Section~\ref{sec:human-exp-ai-assessment}, and designing and testing a tool to bridge the information asymmetry, Section~\ref{sec:design-tool}, led to the definition of a framework to operationalize \acp{AIRS}. This framework represents the potential user experience and interactions of the three main stakeholders in an \ac{AIRS}, having the tool as the main touchpoint.

The phase structure presented in our framework is a synthesis of the work of existing regulatory, institutional, and academic sources that describe the lifecycle of an \ac{AIRS}. The number of phases and naming slightly vary from one another, but the core content remains consistent, consisting in a sequence from participant selection to preparation and testing and ending with the exit and monitoring. The \ac{EU AI Act} \cite{EUAIAct2024} and the draft implementing act for \acp{AIRS}~\cite{EUComDraftAIRS2025} outline application and selection, participation, sandbox plan, written proof and exit report and post-participation. Complementary \ac{EU} initiatives, such as EUSAiR \cite{EUSAiR2025AIEUSAiR} articulate related phases, with a slightly more granular approach: pre-participation, application and selection, preparation, participation, evaluation and exit, post-participation, reporting and monitoring. Academic sources either apply these phases, \cite{Buscemi2026OperationalisingAct}, or converge on a more simplified approach, \cite{Novelli2025GettingAct}, with only the core phases: application, preparation, testing validation and exit. For the purpose of this study, we adopt this consolidated phase structure against which we situate our framework instead of proposing new phases, as our contribution lies in how the \ac{BNA} designed in section \ref{sec:design-tool} can enhance stakeholder interactions and collaboration among them. 

\begin{figure}[!htb]
    \centering
    \includegraphics[width=1\linewidth]{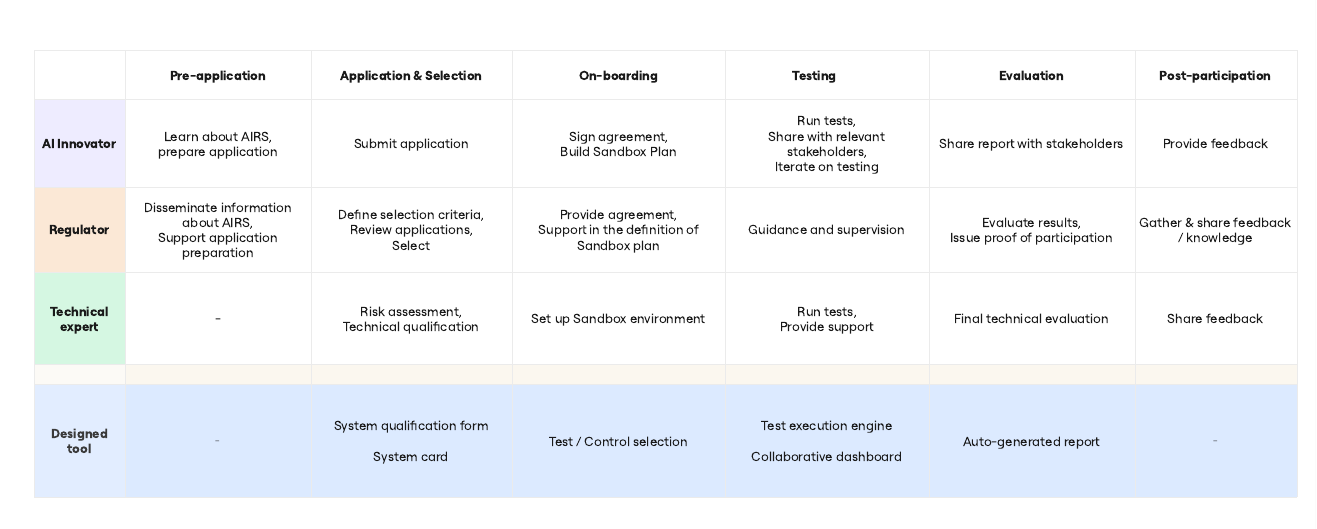}
    \Description{Framework of an AIRS using the tool as the technical backbone. Framework is distributed into 6 phases and each phase explains the tasks of the three stakeholders and the touchpoint of the tool}
    \caption{Framework representing the structured phases of an AIRS engagement, decomposed into stakeholder experiences and tasks. The bottom row contains the role the designed tool plays in support the user experience in each phase.}
    \label{fig:Framework}
\end{figure}

As a result, the framework, shown in Figure~\ref{fig:Framework}, is divided into six general phases, presented in the first row: pre-application, application and selection, on-boarding, testing, evaluation, and post-participation. The designed tool presented in the previous section facilitates the process and iterations of the four core phases: the selection, the on-boarding, the testing and the evaluation, while the framework presents the interactions of three stakeholders in the entire process of the \ac{AIRS}. Each row represents the actions of a stakeholder, while the last row presents how the designed tool fits into the process.
To present the framework, the technical expert and the AI innovator are separated into two different stakeholders each of them with their own actions. However, depending on the organization or specific use case, these two stakeholders could be merged into one.

The first phase of the framework is the \textit{pre-application}. For the AI innovators this phase is especially relevant, as it is in this phase where they will decide to participate in the \ac{AIRS}. They have to learn about what the \ac{AIRS} can do for them, understand the benefits of participating and prepare the application. For this to happen, the \acp{NCA} have to disseminate information and benefits of participating in an \ac{AIRS} to the ecosystem, as well as provide support to the AI innovators in the application preparation. In this particular phase, the technical experts do not play a role yet, however, they can help orient potential users to the \ac{AIRS} and help these users understand the value and benefits of applying. Requirements for a successful pre-application phase include a good information campaign and a harmonized application procedure across \acp{AIRS}. In this phase the designed tool does not play a role yet.

The second phase is the \textit{application and selection} of participants. In this phase the AI innovators submit the application and receive the notification of the application decision from the \ac{NCA}. In order to do this, the \acp{NCA} need to first establish the eligibility and selection criteria, analyze the submitted project against those criteria and select the applicants. During this process, the technical expert can provide support to the \acp{NCA} by providing technical support to perform risk assessment and classification of the submitted project, provide technical advice to identify requirements and feasibility and provide a technical qualification of the AI system under evaluation. While the \ac{NCA} needs to provide a platform for the AI innovators to apply, the technical qualification of the AI system can be performed through the design tool presented in Section~\ref{sec:design-tool}. This will allow AI innovators and \acp{NCA} to download a structured system card, e.g.\, the proposed AI cards~\cite{golpayeganiAICardsApplied2024a}, with the details of their AI system, such as intended purpose, target users and high-level information about data governance, human oversight, transparency to end users, to name a few.

Once the applicants are selected, they move forward to the third phase, \textit{on-boarding}. In this phase the AI innovators will have to sign an agreement for accepting the general terms and conditions of the \ac{AIRS} and complete a sandbox plan in alignment with the \ac{NCA}. Thus, if they have performed any testing prior to the participation in the \ac{AIRS} they will have to provide the corresponding test results. Regarding the \ac{NCA}, in this phase they will have to provide the AI innovators with the agreement and provide support in the creation of the sandbox plan. As for the technical expert, with the help of the tool (Section \ref{sec:design-tool}) and based on the qualification of the AI system made in the previous phase, this phase consists in setting up the technical sandbox environment. This involves the selection of relevant tests, datasets, and controls. The interaction of all three stakeholders in this phase is key to ensure alignment throughout the following phases.

The fourth phase of the framework consists in performing the tests in the tool following the established plan. The testing can be done by both the AI innovator or technical expert, while the \ac{NCA} can provide guidance and support. Through the use of the designed tool, the results of the testing will be shown in a collaborative dashboard, where all the stakeholders can see the relevant information and make comments. The open architecture is designed to be modular, which allows customized dashboarding tools such as PowerBI or Superset increasing the utility of the tool as a \ac{BNA}. This will allow to make all the necessary iterations in the testing process while identifying the strengths, gaps and risks of the AI system.  All the changes made to and iterations of the process will be recorded thanks to the tamper-proof logs, giving a detailed outline of all the performed tests.

Upon conclusion of the testing phase, it is followed by one final evaluation and interpretation of the test results. The tool automatically generates a report, facilitating the reporting phase for all stakeholders. This report contains an overview of the system and the assessment results, with a more detailed analysis of the results of each assessment tool used.
This includes, e.g., highlighting failed test cases for further analysis by technical experts.
The AI innovator will be in charge of sharing this report with the relevant stakeholders. As for the \ac{NCA}, they will have to evaluate said results and provide the AI innovator with a written proof of participation in the \ac{AIRS}. Also, relevant and non-confidential findings will be published by the \ac{NCA}. In this phase the technical expert can also provide a final technical evaluation of the strengths, gaps, and residual risks of the AI system under assessment. 

The final phase of the framework is the post-participation, with the participation in the \ac{AIRS} having formally concluded but communication among the stakeholders still ongoing to sustain their interest and involvement. 
A key action in this phase is to share the generated knowledge and gather feedback. The feedback activity refers to evaluating the process both internally, and with the participants, both AI innovators and technical experts. It is noteworthy that the majority of the processes that engage a variety of stakeholders conclude in the previous phase, without a dedicated phase addressing what occurs once the process is completed, undermining its impact.
Hence, in this phase it is important for the \acp{NCA} to gather the feedback of other stakeholders and share the knowledge gathered trough the process with them.

This framework aims at bridging the gap between the stakeholders when assessing an AI system by providing a common tool adapted to each of their needs. Grounded in the insights presented in Section~\ref{sec:human-exp-ai-assessment}, this tool provides a common platform for both technical and non-technical users to interact regarding the assessment of an AI system. It reduces the complexity on setting up testing environments and \acp{AIRS}.

\section{Discussion}
This paper explores how AI innovators, technical experts, and regulators experience the collaboration on \acp{AIRS}, what design knowledge emerges from building a tool to mediate that collaboration, and how that knowledge can inform the operationalization of \acp{AIRS} more broadly. Put together, our findings suggest that the central issue for a well-functioning \ac{AIRS} is not the deficit of information, but the sense-making of it all, as stakeholders hold substantial and legitimate expertise, but lack a shared understanding. This is relevant, as it re-frames the design challenge of an \ac{AIRS}. Simply supplying more documentation or guidance is not enough for its proper operationalization. There is a need for artifacts that will close the information asymmetry among the stakeholders, offering a means that can be used to negotiate meaning. This extends the view of sandboxes articulated by \citeauthor{guioespanolRegulatorySandboxesAI2025} \cite{guioespanolRegulatorySandboxesAI2025}, understanding \acp{AIRS} as fundamentally learning environments, with their central design problem being a communicative one.

The three design iterations presented in Section\ref{sec:design-tool} address that challenge, giving an empirical account of \acp{BNA} \cite{Deckenbrunnen2026BathtubsUncertainty} in practice. Each aspect of the tool, such as the qualification of the system, the harmonized tests and controls,  the dashboard, or the report, all perform a distinct sense-making function, providing a common ground for the different stakeholders involved. Putting these artifacts together, across the \ac{AIRS} life-cycle, they function as the backbone that the framework presented in Section~\ref{sec:framework} formalizes. This expands on the \ac{BNA} concept for a regulatory context specifically. Instead of an artifact bridging multiple communities of practice, it works as a collection of touchpoints, each working at a specific phase of the process, whose cumulative effect is the reduction of information asymmetry across the whole \ac{AIRS} life-cycle.

This finding lies within a long standing conversation in the HCI community on designing for collaboration across diverse expertise. 
HCI has previously used boundary objects and shared vocabularies as a way of supporting multi-stakeholder collaboration, such as using the notion of \acp{BNA} as a way to coordinate data practices across stakeholders \cite{Lee2026EmbracingResponse}. 
This underscores a broader turn in the field towards using \acp{BNA} to hold high-stakes multi-stakeholder discussions under uncertainty or in the absence of established consensus.
Our work contributes to this turn by expanding it into a regulatory context, where participants are multi-disciplinary and structurally different. 
AI innovators, regulators, and technical experts hold different levels of authority, accountability and literacy over the same system, yet they all need to interpret the same results.
Designing for common ground goes beyond translation between vocabularies, as it needs to do so without undermining the legitimate expertise that each stakeholder brings to the exchange.
In this sense, this work extends the multi-stakeholder design practices of HCI toward the emerging field of \acp{AIRS}.

Another relevant finding is the direction of the design process itself. The feedback gathered on the first iteration showed that the tool was well understood by technical experts, while not as easily understood by other less technical users. This issue reflects, on a small scale, what the broader literature for responsible AI suggests, as organizations building AI tend to default to efficiency and technology-first practices, often neglecting the user needs \cite{Rakova2021WhereReality}. 
Within an \ac{AIRS} this can create tension among stakeholders, as the efficiency of a technical tool might reduce the burden for the AI innovators and technical experts but be inefficient to share with other stakeholders like the \ac{NCA}.
By applying the \ac{RtD} methodology, participants were able to engage with the designed system in every iteration, providing insights that only became visible while observing stakeholders' behavior when interacting with the prototype.
This trajectory reflects the importance of humans in the design process as designing an \ac{AIRS} relies on the shared understanding across multiple expertises, highlighting the relevance of stakeholder involvement in the process.

These findings converge in the presented framework. Situating the tool as the backbone of the framework and giving each stakeholder a specific set of actions in each step, the framework provides a sense-making mechanism for \acp{AIRS} bridging the information asymmetry. While the \ac{EU AI Act} \cite{EUAIAct2024} and the draft implementing act for \acp{AIRS}~\cite{EUComDraftAIRS2025} specify that a \ac{NCA} must supervise and an innovator must assess its AI system, this framework specifies \textit{how} that supervision, planning and testing can be mediated so that the resulting exchange is meaningful to all stakeholders. In this sense, the framework's contribution offers a shared artifact that helps in the operationalization of \acp{AIRS}, providing a common ground for all stakeholders.

The implications of the proposed solution extend beyond the use case of this paper.
As outlined in Section \ref{sec:AIRS-legal} the \ac{EU AI Act} mandates that every member state establishes at least one \ac{AIRS}, yet leaves considerable discretion to \acp{NCA} regarding its operationalization. This discretion risks fragmented or incompatible approaches that could undermine a consistent interpretation of the act. 
This risk, anticipated by the Commission itself, is reflected in the draft implementing act~\cite{EUComDraftAIRS2025}.
However, this harmonization attempt operates at the level a legal procedure, without focusing on the day-to-day multi-stakeholder experience of a \ac{AIRS}.
The framework operationalized by the technical \ac{BNA} proposed in this paper addresses that gap. It formalizes a common structure of touchpoints and stakeholder exchanges that \acp{NCA} across different member states can adopt as a practical layer under the guidelines of the \ac{EU AI Act}. 
This offers a replicable structure for setting up their own \ac{AIRS} without the need of designing the collaboration tools from scratch.
For AI innovators, on the other hand, it offers a more consistent point of reference across sandboxes, reducing the cost of learning new collaboration models each time they engage with a different \ac{NCA}.

Beyond the \ac{EU}, the relevance of the framework lies in providing a generic structure of multi-stakeholder collaboration in \acp{AIRS}. \acp{AIRS} are being established well beyond the \ac{EU} context, all varying in institutional maturity \cite{guioespanolRegulatorySandboxesAI2025}.
Since the framework's touchpoints, such as harmonized tests or reporting are organized around the sense-making needs of innovators, technical experts, and regulators rather than around any single legal text, they are, in principle, transferable to other legislative systems that face a similar need to structure trust and communication between stakeholders under legal uncertainty.
That said, this transferability remains a claim as differences in institutional capacity and regulatory culture outside the EU may require some modifications to the framework.

Finally, our findings are grounded in a single, evolving tool developed and evaluated primarily through semi-structured feedback from stakeholders rather than sustained longitudinal use within a live, operating \ac{AIRS}. 
Therefore, the framework should be read as a design proposal validated through iterative exposure to stakeholders.
Additionally, confidentiality agreements with participating organizations mean that raw quotes, case-specific details, and organization-level findings cannot be shared. All reported findings are consequently presented at the level of recurring patterns across sessions rather than as directly attributable statements.

\section{Conclusion}
This paper examined how \acp{AIRS} can be designed to support genuine stakeholder collaboration, with special focus on AI innovators, technical experts, and regulators, moving beyond merely procedural requirements. Through a \ac{RtD} process grounded in workshops and three iterations of a digital tool, we concluded that the core challenge these stakeholders face is not a lack of information but a lack of shared sense-making across differing vocabularies and expertise. Building on this finding we created a \ac{BNA} that bridges the asymmetry across the \ac{AIRS} life-cycle providing an artifact that enhances the experience of the stakeholders. Therefore, as \acp{AIRS} multiply across \ac{EU} member states, we hope this framework offers a path towards the kind of interoperable and human-centered practice.
Future work should build on these findings through user studies within a real, operating \ac{AIRS}, observing how AI innovators, technical experts, and regulators engage with the tool and framework under the full conditions of a live sandbox.
A second direction is to pilot the framework and tool across \acp{AIRS} in different EU member states, testing directly the interoperability and replicability, examining how its structure holds up against the procedural variation between \acp{NCA}.

\appendix
\section*{Acknowledgments}
We thank the Luxembourg AI Factory (L-AIF, grant agreement ID 101234366) for supporting this work and all the stakeholders that participated in the design process.

\printbibliography

\end{document}